\documentclass[preprint, journal]{vgtc}       

\ifpdf
  \pdfoutput=1\relax                   
  \usepackage{graphicx}                
  \DeclareGraphicsExtensions{.pdf,.png,.jpg,.jpeg} 
\else
  \usepackage{graphicx}                
  \DeclareGraphicsExtensions{.eps}     
\fi%

\graphicspath{{figures/}{pictures/}{images/}{./}} 

\usepackage{microtype}                 
\PassOptionsToPackage{warn}{textcomp}  
\usepackage{textcomp}                  
\usepackage{mathptmx}                  
\usepackage{times}                     
\usepackage{cite}                      
\usepackage{tabu}                      
\usepackage{booktabs}                  

\usepackage{comment}
\usepackage{graphicx}
\usepackage{url}
\usepackage{subfig}
\usepackage{enumitem}

\usepackage[dvipsnames]{xcolor}
\newcommand{\attr}[1]{{\rmfamily \scshape {#1}}}
\newcommand{\ctrl}{{\rmfamily \scshape ctrl}}
\newcommand{\summ}{{\rmfamily \scshape sum}}
\newcommand{\rt}{{\rmfamily \scshape rt}}
\newcommand{\rtsum}{{\rmfamily \scshape rt+sum}}

\newcommand{\emily}[1]{\textcolor{magenta}{#1}}
\newcommand{\alex}[1]{\textcolor{green}{#1}}
\newcommand{\arpit}[1]{\textcolor{brown}{#1}}
\newcommand{\adam}[1]{\textcolor{orange}{#1}}
\newcommand{\delete}[1]{\textcolor{red}{#1}}
\newcommand{\revised}[1]{\textcolor{black}{#1}}
\newcommand{\todo}[1]{\textcolor{olive}{#1}}

\MakeRobust{\emily}
\MakeRobust{\alex}
\MakeRobust{\arpit}
\MakeRobust{\adam}
\MakeRobust{\delete}
\MakeRobust{\revised}
\MakeRobust{\todo}

\renewcommand{\delete}[1]{\unskip}

\ieeedoi{10.1109/TVCG.2021.3114862}

\onlineid{0}

\vgtccategory{Research}
\vgtcpapertype{please specify}

\title{Left, Right, and Gender: Exploring Interaction Traces to Mitigate Human Biases}

\author{Emily Wall*, Arpit Narechania*, Adam Coscia, Jamal Paden, and Alex Endert}
\authorfooter{
\item
 Emily Wall is with Emory University. E-mail: emily.wall@emory.edu.
\item
 Arpit Narechania, Adam Coscia, Jamal Paden, and Alex Endert are with Georgia Tech. E-mail: \{arpitnarechania, acoscia6, jpaden, endert\}@gatech.edu.
 \item
 *Authors contributed equally.
 }

\shortauthortitle{Wall \MakeLowercase{\textit{et al.}}: Exploring Interaction Traces to Mitigate Social Biases}

\abstract{Human biases impact the way people analyze data and make decisions. 
Recent work has shown that some visualization designs can better support cognitive processes and mitigate cognitive biases (i.e., errors that occur due to the use of mental ``shortcuts''). 
In this work, we explore how visualizing a user's interaction history (i.e., which data points and attributes a user has interacted with) can be used to mitigate potential biases that drive decision making by promoting conscious reflection of one's analysis process.
Given an interactive scatterplot-based visualization tool, we showed interaction history in \textit{real-time} while exploring data (by coloring points in the scatterplot that the user has interacted with), and in a \textit{summative} format after a decision has been made (by comparing the distribution of user interactions to the underlying distribution of the data).
We conducted a series of in-lab experiments and a crowd-sourced experiment to evaluate the effectiveness of interaction history interventions toward mitigating bias. 
We contextualized this work in a political scenario in which participants were instructed to choose a committee of 10 fictitious politicians to review a recent bill passed in the U.S. state of Georgia banning abortion after 6 weeks, where things like gender bias or political party bias may drive one's analysis process.
We demonstrate the generalizability of this approach by evaluating a second decision making scenario related to movies. 
\revised{Our results are inconclusive for the effectiveness of interaction history (henceforth referred to as \emph{interaction traces}) toward mitigating biased decision making. 
However, we find some mixed support that interaction traces, particularly in a summative format, can increase awareness of potential unconscious biases.}
} 


\CCScatlist{ 
 \CCScat{K.6.1}{Management of Computing and Information Systems}%
{Project and People Management}{Life Cycle};
 \CCScat{K.7.m}{The Computing Profession}{Miscellaneous}{Ethics}
}

\vgtcinsertpkg

\begin{document}


\firstsection{Introduction}

\maketitle

As the sheer volume and ubiquity of data increases, data analysis and decision making are increasingly taking place within digital environments, where humans and machines collaborate and coordinate to inform outcomes, facilitated by interactive visual representations of data. 
These environments provide a new way to measure and characterize cognitive processes: by analyzing users' interactions with data during use. 
Analyzing user interactions can illuminate many aspects about the user and their process, including identifying personality traits~\cite{Brown2014}, recovering a user's reasoning process~\cite{Dou2009}, and most relevant to the present work, characterizing human biases~\cite{WallBias}.
In this work, we explore how showing a user prior interaction history might be used to \textbf{mitigate potential biases} that may be driving one's data analysis and decision making.

We utilize the technique of \textbf{interaction traces}, a form of provenance~\cite{North2011} visualization in which a user's own previous interactions with the data influence the visual representations in the interface. 
We show interaction traces in two ways: \emph{in-situ} interaction traces alter the color of visited data points in a scatterplot based on the frequency of prior interactions (Figure~\ref{fig:study_summary}E), and \emph{ex-situ} interaction traces are shown in an additional view of the data that compares the distribution of a user's interactions to the underlying distributions in the data (Figure~\ref{fig:study_summary}F).

\revised{We operationalize \textbf{biased behavior} as deviation from a baseline of equally probable interactions with any data point. 
It can be conceptualized as a model mechanism~\cite{wall2018four}, captured using bias metrics~\cite{WallBias}, and may correspond to other notions of societal or cognitive bias.
Similarly then, a \textbf{biased decision} is one which reflects choices that are not proportional to the data.
This definition of bias serves as a point of comparison for user behavior and decision making, but, as described in~\cite{WallBias}, is not inherently negative and requires interpretation in context by the user given their goals. 
We posit that visualization of interaction traces will lead to reflection on behavior and decision making, increasing awareness of potential biases.}
Importantly then, our definition of \textbf{bias mitigation} is a reduction in \emph{unconscious} biases, which we aim to address by promoting user reflection~\cite{sengers2005reflective} about factors driving their decision making processes. 
In particular, we examine the effectiveness of visualizing traces of users' interactions, where effectiveness is measured by (1) \textit{behavioral changes}, (2) \textit{changes in decisions made}, and (3) \textit{increased cognitive awareness}. 

To assess the impact of interaction traces toward mitigating potential biases, we designed an interactive scatterplot-based visualization system (Figure~\ref{fig:study_summary}). 
We conducted a crowd-sourced experiment in which users performed two decision making tasks in the domains of (1) politics and (2) movies. 
In the political scenario, we curated a dataset of fictitious politicians in the U.S. state of Georgia and asked participants to select a committee of 10 responsible for reviewing public opinion about the recently passed Georgia House Bill 481 (Georgia HB481), banning abortion in the state after 6 weeks. 
In this scenario, several types of bias may have impacted analysis, including gender bias (i.e., bias favoring one gender over another), political party bias (i.e., voting along political party lines, regardless of potential ideological alignment from candidates in another party), age bias (i.e., preferential treatment of candidates based on age), and so on. 
Participants in the experiment also completed a parallel task in the domain of movies: to select 10 representative movies from a dataset of similar size and composition. 
In this task, we anticipated that participants' decisions would be driven by idiosyncrasies of their individual preferences.

For the given tasks, we assessed four interface variations: \ctrl, \summ, \rt, and \rtsum. 
The \ctrl{} interface served as the control system, which we compared to variations that provided either \textit{real-time} 
(\rt) or \textit{summative} (\summ) views of the user's interaction traces (or both, \rtsum). 
\revised{Our experiments yielded mixed results, offering support that interaction traces, particularly in a summative format, can lead to behavioral changes or increased awareness, but not substantial changes to final decisions. 
Interestingly, we find that increased awareness of unconscious biases may lead to amplification of individuals' conscious, intentional biases.}
We emphasize that regardless of domain, our goal is not to address overt biases (e.g., in the form of discrimination) in this work; rather, we believe visualization \emph{can} have an impact on increasing user awareness of potential unconscious biases that may impact decision making in critical ways. 

In this work, we highlight the following contributions: 
\begin{enumerate}[nosep]
\item We utilize a technique for showing interaction history, (referred to as \emph{interaction traces}, Section~\ref{sec:system}), 
\item We present results of three formative in-lab studies that describe exploratory and qualitative findings (Section~\ref{sec:formative}), and 
\item We present results of a crowd-sourced study that describes quantitative effects of interaction traces (Section~\ref{sec:results}).
\end{enumerate}
In the following sections, we present a description of the datasets and interface used in the studies, findings from the in-lab and crowd-sourced experiments, and a discussion of how these results can inform the design of future systems that can mitigate potentially biased analyses.

\section{Related Work}


\revised{Wall et al. introduced four definitions of the term ``bias'' in data visualization, relating to human cognitive, perceptual, and societal biases, and a fourth usage as a \emph{model mechanism}~\cite{wall2018four}.
We adopt the fourth perspective.}
Namely, we utilize computational metrics to characterize how a person's interactive behavior deviates from a baseline model of expected behavior~\cite{WallBias}. 
Specifically, we model and visualize how a user's interaction sequences deviate from uniform behavior. \delete{to promote conscious reflection of factors that drive one's analysis process.}
\revised{This model serves as a benchmark against which a user can compare, interpret, and reflect on their behavior.
We intend this usage to have a neutral connotation -- deviation from a baseline is neither good nor bad, but relies on a user's interpretation of the metrics in context.}
In a political scenario (one task in our experiment), these metrics can be used to indicate when a user's attention is skewed toward e.g., a particular political party, politicians' genders or ages, etc. 
\delete{In a scenario about movies, these metrics can characterize a user's analytic focus on attributes such as movie genre or critic ratings e.g., via Rotten Tomatoes.}

\revised{These metrics capture deviations which may correspond to systematic biases, e.g., cognitive or societal, which inherently impact the lens through which a person analyzes and makes decisions from data.}
\delete{Decision making may be impacted by a multitude of human biases, including social, cognitive, and perceptual biases.  
This work addresses the former two.}
In Cognitive Science, bias can describe an irrational error that results from heuristic decision making~\cite{Kahneman2005,Tversky1974,kahneman2011thinking}.
Alternatively, it can refer to a rational decision made under certain constraints (e.g., limited time or high cognitive load)~\cite{Gigerenzer2011,Gigerenzer2009,gigerenzer2004fast}.
Cognitive biases can thus influence how people make decisions when ``fast and frugal'' heuristics~\cite{gigerenzer2004fast} are employed in place of concerted, deliberative thinking~\cite{evans2013dual}.

In Social Sciences, bias often refers to prejudices or stereotypes that are relevant in society (e.g., racial bias or gender bias). 
In this work, we refer to such biases as \emph{social biases}. 
These biases can have far-reaching impacts, such as propagating racial or gender bias to machine learning~\cite{garg2018word,manzini2019black}.
 
Social biases may be influenced by cultural norms, individual experiences or personality variations, and they can shape our decision making in a conscious or an \delete{implicit} \revised{unconscious} manner~\cite{greenwald2006implicit}. 
These biases can have severe implications in a variety of decision making domains. 
For example, consider the impact of racial bias in hiring. 
Researchers have found discrimination, either conscious or \delete{implicit} \revised{unconscious}, based on racial name trends~\cite{bertrand2004emily}, showing that equivalent resumes with traditionally White names receive 50\% more callbacks from job applications than resumes with traditionally African American names. 
As a result, companies may lack a diverse workforce, which can have implications on employee turnover, group isolation or cohesion, workplace stress, and so on~\cite{reskin1999determinants}. 

In the visualization community, bias has garnered increasing attention.
Researchers have cataloged relevant biases~\cite{dimara2018task} and proposed methods for detecting the presence of a particular type of bias~\cite{WallBias,WallFormative,gotz2016adaptive,cho2017,dimara2017attraction,valdez2018priming}.
Other recent works proposed or categorized methods for mitigating bias~\cite{dimara2019mitigating,sukumar2018visualization,law2018designing,WallDesignSpace}. 
Within Wall et al.'s design space of bias mitigation techniques for visualizations~\cite{WallDesignSpace}, our proposed system manipulates the visual representation to show metrics about a user's analysis in a minimally intrusive, orienting~\cite{ceneda2017characterizing} fashion, to ultimately facilitate more balanced decision making. 
Distinct from prior work on bias mitigation in visualization, \revised{we focus on increasing awareness of \emph{unconscious} biases which could correspond to cognitive or social biases,} including gender bias and political bias (e.g., bias towards one political party), among others. \delete{with the goal of mitigating these during decision making.}

To mitigate potential biases driving decision making, we are motivated by literature in Cognitive Science on nudging~\cite{thaler2009nudge} and boosting~\cite{grune2016nudge}, that can influence people's \textit{behavior} and \textit{decision making} by altering the choice architecture (i.e., the way that choices are presented) or improving individuals' decision making competences.
We apply this analogy in the context of visualization with the goal of ``nudging'' users toward a less biased analysis process.
In visualization research, prior work has shown some ability to impact user behavior, resulting in more broad exploration of the data (e.g., by coloring visited data points differently ~\cite{feng2017hindsight} or by adding widgets that encode prior interactions~\cite{willett2007scented}).
Furthermore, we are inspired by work on reflective design~\cite{sengers2005reflective}, wherein our purpose is not to prescribe an optimal decision to users, but rather to encourage thoughtful reflection on motivating factors of those decisions while users maintain full agency.
We describe the visualization system and interaction traces in Section~\ref{sec:system}.

\section{Bias Metric Review}
\label{sec:bias_metric_review}
While several metrics have been proposed to quantify aspects of a user's analysis process (e.g.,~\cite{feng2019patterns,jankun2000spreadsheet,monadjemi2020competing,WallBias}), here we focus on bias metrics introduced by Wall et al.~\cite{WallBias} which are theoretically applicable to various types of bias and have been used for initial characterization of anchoring bias~\cite{WallFormative}. 
We quantify bias using the data point distribution (DPD) and attribute distribution (AD) metrics~\cite{WallBias}. 
These metrics characterize, along a scale from 0 (no bias) to 1 (high bias), how a user's interactive behavior deviates from expected behavior. 
In this case, expected behavior is defined by equal probability of interaction with any given data point in the dataset. 

Consider a dataset of politicians.
Data point distribution (DPD) describes how the user's interactions are distributed over the points (politicians) in the dataset. 
Uniform interactions over all politicians will result in a low metric value (less biased), while repeated interaction with a subset of the data (e.g., only Republicans) will result in a higher metric value (more biased).

Attribute distribution (AD) considers how the users' interactions across the data map to the underlying distributions of each attribute. 
That is, if the dataset has politicians with an average Political \attr{Experience} of 9 years, but the user focuses almost exclusively on politicians with 15+ years of \attr{Experience} (potentially revisiting the same subset of experienced politicians), the attribute distribution metric for \attr{Experience} would be high (more biased). 
Alternatively, if the user's interactions are proportional to the dataset, the metric value would be low (less biased).
See ~\cite{WallBias} for the precise formulation of the bias metrics.
These metrics drive the visualization design in this paper that shows a user's interaction traces as they make their decisions.

\section{Methodology}
\label{sec:methodology}

\begin{figure*}[hbtp]
\centering
\makebox[\textwidth]{\includegraphics[width=.85\paperwidth]{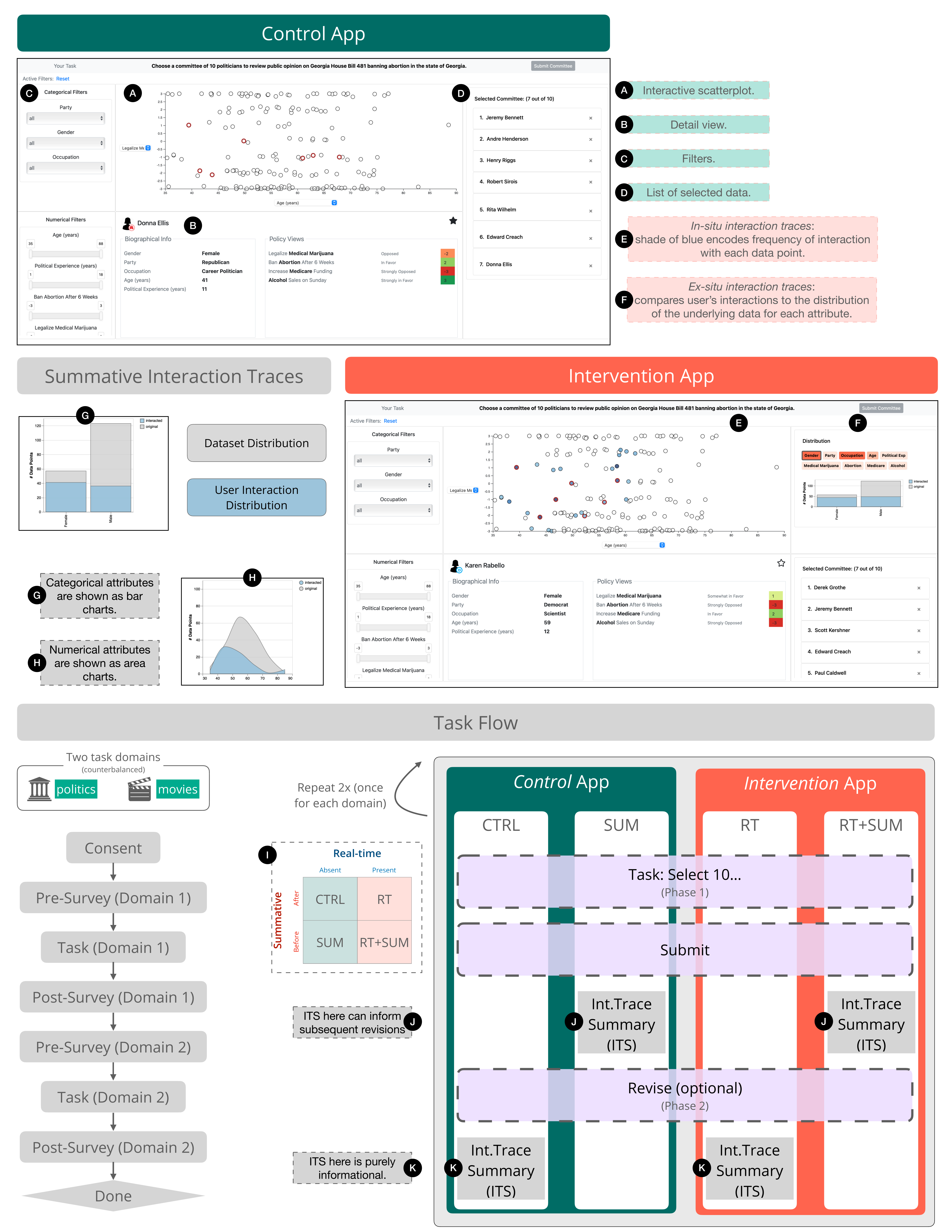}}
\caption{A summary of the interfaces and procedure for this experiment.}
\label{fig:study_summary} 
\vspace{-1em}
\end{figure*}

To study the effect of visualizing interaction traces toward mitigating bias, we conducted a series of in-lab studies and a crowd-sourced experiment to test four interface variations (\ctrl, \summ, \rt, \rtsum). 
In this section, we describe two tasks and datasets in the domains of politics and movies (Section~\ref{sec:task}) and the implementation of a visualization system that realizes interaction traces to serve as the testbed for subsequent experiments (Section~\ref{sec:system}). 

\subsection{Tasks \& Datasets}
\label{sec:task}

We selected two complementary tasks (counterbalanced within subjects) to observe how people would respond to interaction traces in the presence of a variety of potential biases, described below for each task. 

\subsubsection{Politics}
\label{sec:task_politics}
\noindent\textbf{Task. }
The USA has a two-party political system: Democrats and Republicans~\cite{aldrich1995parties}. 
In Georgia's General Assembly, committees may be formed to explore complex issues, draft legislation, and make recommendations~\cite{gaCongress}.
Many such committees, particularly subcommittees focused on specific issues, may be formed by top-down appointment~\cite{gaCongress}. 
With membership in committees often decided by an individual or by few, the decision can be subject to an individual's biases.

In May 2019, Georgia's incumbent Governor Brian Kemp signed Georgia House Bill 481 (Georgia HB481) banning abortion after 6 weeks (earlier than the previous state law of 20 weeks)~\cite{romo_2019}. 
Scheduled to take effect in January 2020, the bill was received by the public with significant controversy~\footnote{A federal judge permanently blocked Georgia HB481 in July 2020, finding it in violation of the U.S. Constitution~\cite{gpbInjunction}}. 
Supporters of Georgia HB481 (colloquially referred to as a ``Heartbeat Bill'') hoped it would lead to overturning of Roe v. Wade, 410 U.S. 113 (US federal court decision protecting a woman's right to an abortion, 1973), while opponents hoped to challenge the bill before it became law. 

Given a dataset of fictitious politicians, participants were given the following task: \textit{Imagine you are engaged in political decision-making in the state of Georgia. The debate about abortion is ongoing, with variations of these bills cropping up across other states in the United States, which can potentially learn from the ongoing debate in Georgia. Select a committee of 10 candidates that you feel should review public opinion in Georgia on the controversial Georgia HB481}.
We selected this task to simulate a realistic decision making scenario in American politics and evaluate our interventions in a politically and socially relevant context. 
Furthermore, this topic and dataset can elicit \revised{a number of factors that may influence an individual's decision making process, including personal preferences as well as multiple} types of social biases (e.g., gender bias or political party bias), both \delete{explicit} \revised{conscious} and \delete{implicit} \revised{unconscious}. 

\medskip
\noindent\textbf{Dataset. }
We generated a dataset of 180 fictitious politicians, representing the composition of the Georgia General Assembly~\cite{gaGeneralAssembly}. 
Each row in the dataset represents a politician, described by the following attributes: \attr{Gender}, \attr{Political Party}, \attr{Occupation}, \attr{Age}, and \attr{Experience}, along with numerical representations $\in[-3,3]$ of the politician's view on topics such as \attr{Banning Abortion After 6 Weeks}, \attr{Legalizing Medical Marijuana}, \attr{Increasing Medicare Funding}, and \attr{Banning Alcohol Sales on Sundays} (positive numbers indicate that the politician is in favor, while negative numbers indicate that the politician is opposed). 
Politicians' names are artificially generated from US census data~\cite{nameGeneratorRef}.
The dataset contains 59\% Republicans, of which 14\% are female; and 41\% Democrats, of which 57\% are female, mimicking the distributions in the Georgia General Assembly~\cite{womenStateLegislatures}.
The ages, political experience, and occupations were derived from data on the 115th U.S. House of Representatives~\cite{congressRef}. 
The policy views were generated to represent general party voting trends (e.g., Democrats tend to be opposed to banning abortion, while Republicans tend to be in favor of the ban) with the strength of those views representing recent increasing polarity~\cite{ringel2019right} in the USA political system (e.g., fewer politicians have neutral positions or positions against the party trend).

\subsubsection{Movies}
\noindent\textbf{Task. }
Given a dataset of fictitious movies, participants were given the following task: \textit{Analyze the data to pick 10 movies that you feel represent the collection of movies in the dataset as a whole.}
We selected this task to complement the political scenario. 
It represents a parallel task (selecting a representative subset) in a domain that the general public is familiar with (movies). 
We hypothesize that this task may elicit an entirely different set of (less obviously dangerous) biases, based on idiosyncrasies in one's movie preferences.
For instance, participants may make selections for movies by focusing on attributes of the data that are most familiar to them (e.g., \attr{Rotten Tomatoes Rating}) while disregarding others that have a lesser impact on their own movie habits (e.g., \attr{Running Time}).
The instructions for both tasks were intentionally vague to avoid suggesting any particular criteria for selecting politicians / movies.

\medskip
\noindent\textbf{Dataset. }
The movies dataset was adapted~\cite{moviesDataset} to match the general structure of the political dataset. 
We sampled 180 movies from the dataset and selected 9 attributes in total (3 categorical, 6 numerical): \attr{Content Rating}, \attr{Genre}, \attr{Creative Type}, \attr{Worldwide Gross}, \attr{Production Budget}, \attr{Release Year}, \attr{Running Time}, \attr{Rotten Tomatoes Rating}, and \attr{IMDB Rating} to match the dimensionality of the political dataset.
In pilot studies, we found that (1) real movie titles were problematic because participants relied heavily on familiarity of titles rather than the data before them; and (2) anonymized identifiers (e.g., ``Movie1'', ``Movie2'', ...) led participants to be less engaged with the task.
For consistency with the political scenario, we generated fictitious movie titles~\footnote{\url{https://thestoryshack.com/tools/movie-title-generator/}} so that participants would be more engaged with the task while not relying only on familiarity of titles. 
Complete datasets and analyses are included in supplemental materials\footnote{\label{foot_supplemental}
{\url{https://github.com/gtvalab/bias-mitigation-supplemental}}}.

\subsection{System}
\label{sec:system}

\noindent\textbf{Overview. }
For our experiments, we utilized a simplified version of Lumos~\cite{Lumos}, a visualization system to support data exploration while promoting reflection and awareness during visual data analysis. 
To assess the effectiveness of visualizing interaction traces, we produced two versions of the visualization system: a Control version of the interface, and an Intervention version of the interface, which was modified to visualize traces of the user's interactions with the data in real-time (Figure~\ref{fig:study_summary}). 
Components A-D in Figure~\ref{fig:study_summary} are common across the Control and Intervention interfaces. 
The primary view is an interactive scatterplot (A), where the x- and y-axes can be set to represent attributes of the data via selection in a drop-down menu. 
Hovering on a point (politician / movie) in the scatterplot populates the detail view (B), which shows all of the attributes of that data point. 
Filters for categorical (e.g., \attr{Gender}, \attr{Occupation}, etc. in the political dataset; \attr{Genre}, \attr{Content Rating}, etc. in the movies dataset) and ordinal \& numerical attributes (e.g., \attr{Age}, \attr{Experience}, etc. in the political dataset; \attr{Running Time}, \attr{IMDB Rating}, etc. in the movies dataset) can be adjusted on the left-hand side of the interface (C) using drop-down menus and range sliders. 
Clicking on the point in the scatterplot or on the star icon in the detail view adds the politician / movie to the selected list (D). 
Selected data points are shown in the scatterplot with a thick red border.

\smallskip
\noindent\textbf{Interaction Traces. }
In the Intervention interface, user interaction traces are shown in \emph{real-time} in the interface with respect to \textit{data points} and with respect to \textit{attributes}. 
First, the points in the scatterplot are given a blue fill color (in-situ interaction traces) once the user has interacted with the data point, with darker shades representing a greater number of interactions (DPD metric~\cite{WallBias}; Figure~\ref{fig:study_summary}E). 
The Control interface, by comparison, uses no fill color on the points (Figure~\ref{fig:study_summary}A).
Second, the top right view (Figure~\ref{fig:study_summary}F) compares the user's interactions to the underlying distributions of the data for each attribute (ex-situ interaction traces). 
The attribute tags are colored with a darker orange background when the user's interactions deviate more from the underlying data and with a lighter orange or white background when the user's interactions more closely match the underlying distribution of data (AD metric~\cite{WallBias}). 
Categorical attributes (\attr{Gender} pictured) compare user interactions to the underlying dataset using bar charts, where gray represents the underlying distribution of data (approximately 32\% women, 68\% men) and a superimposed blue bar represents the distribution of the user's interactions (approximately evenly split between women and men).
Numerical attributes compare user interactions to the underlying data distributions using area curves. 

\smallskip
\noindent\textbf{Real-Time v. Summative. }
The interaction traces pictured in Figure~\ref{fig:study_summary}(E-F) in the Intervention interface are shown in \emph{real-time}.
We also show interaction traces in a \emph{summative} format, depicted in Figure~\ref{fig:study_summary}(G-H), after the user has made a decision (choosing 10 politicians or 10 movies). 
We hypothesize that both real-time and summative formats may be beneficial in different ways. 
In real-time, interaction traces may help users maintain awareness throughout their analysis process about the distribution of their analytic focus across the data. 
In a summative format, interaction traces may be easier to process and adjust from in subsequent analyses without the additional simultaneous cognitive load of the decision itself. 
We test variations of both in our experiment.

\section{Formative In-Lab Study Results}
\label{sec:formative}


We conducted three (3) formative in-lab studies, described in turn below. 
These formative studies utilized a similar task as Section~\ref{sec:task_politics} about political decision making along with earlier variants of the same control and intervention interfaces, described in Section~\ref{sec:system}.
Analysis from these formative studies was largely qualitative and exploratory~\cite{tukey1980we} in nature, informing the hypotheses and design of the confirmatory crowd-sourced experiment described in Section~\ref{sec:results}.

\subsection{In-Lab Study 1}
In the first formative study, 6 participants utilized the Control interface to choose a political committee. 
Our goal was to observe a baseline of user behavior and choices. 
Many participants intentionally balanced their political committee along several attributes (seeking ``balanced representation'' -- P02). 
For example, four participants balanced by \attr{Gender} (5 men and 5 women). 
The same four also balanced by \attr{Party} (5 Republicans and 5 Democrats). 

The ways that participants \emph{biased} their committee selections were explicit but nuanced. 
For instance, while P05 balanced across \attr{Gender} and \attr{Party}, they ultimately chose a committee with all 10 members opposed to the bill, explicitly prioritizing ``members (who) were very opposed to the bill.'' 
We generally observed that \textbf{participants were able to \emph{maintain awareness} about potential biases driving their decision making}, which we hypothesized was the result of the relatively small version of the political dataset used in this study (144 data points and 5 attributes). 
Subsequent formative studies increased data dimensionality, from which we observed greater difficulty in maintaining conscious bookkeeping of attributes that impacted decision making.

\subsection{In-Lab Study 2}
\begin{figure}[t]
   \centering
   \includegraphics[width=\columnwidth]{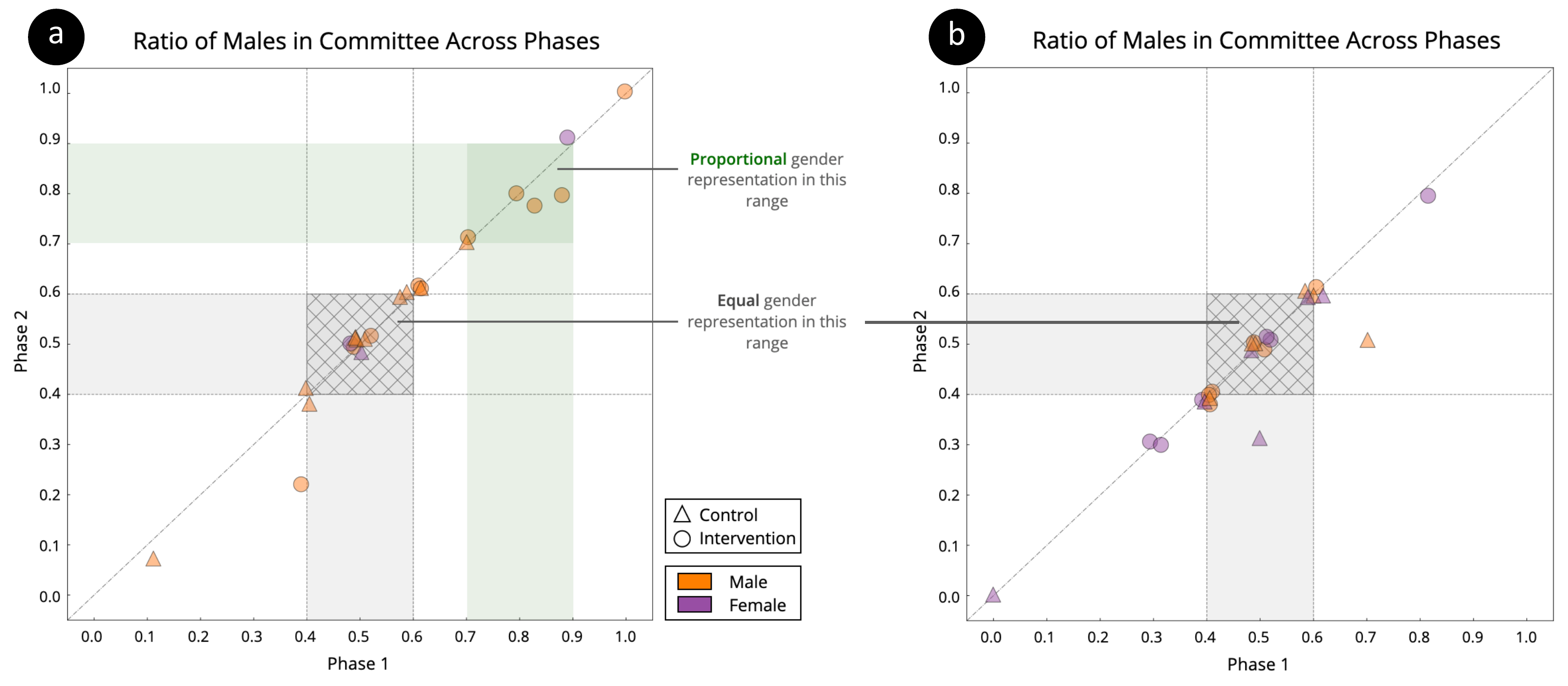}
   \caption{\attr{Gender} balance in committees chosen by 24 participants in (a) Formative Study 2 and (b) Formative Study 3. Balance is shown as the ratio of men in each participant's committee in Phase 1 (x-axis) and Phase 2 (y-axis) (shape encodes condition; color encodes participant gender).}
   \label{fig:ratios}
   \vspace{-1em}
\end{figure}

In the second formative study, 12 participants each utilized the Control and Intervention interfaces to choose a political committee (24 participants in total). 
Our goal was to observe the effects of interaction traces on users' behavior and subsequent decisions.
Exploratory analyses revealed some notable differences between participants' behavior who used the Control v. Intervention interface. 
In particular, for the \attr{Age} attribute, Control participants tended to have higher Attribute Distribution (AD) bias metric values over time than Intervention participants, suggesting that \textbf{Intervention participants interacted with politicians whose ages were more proportional to the underlying dataset than Control participants} ($\mu_C = 0.857$, $\mu_I = 0.729$, $H = 3.360$, $p = 0.057$). 
Furthermore, participants who saw interaction traces \textbf{(Intervention) trended toward choosing more \emph{proportional} gender composition of committees in the political task} (Figure~\ref{fig:ratios}a); however, this trend was not replicated in the third and final formative study (Figure~\ref{fig:ratios}b), potentially due to the introduction of confounding factors, described later.

We also observed instances where interaction traces may have led to altered behavior. 
For instance, after interacting with the interaction trace view, one participant's bias toward \attr{Party} sharply decreased (as observed by the AD bias metric). 
One possible explanation is that the user observed bias in their interactions toward Democratic politicians in the interaction trace visualization and consequently went on to focus on Republicans to reduce the bias.

As captured by Likert ratings, participants found the \textit{summative} metric visualization (4.5 / 5) more useful than \textit{real-time} (4 / 5 for in-situ and 3 / 5 and 4 / 5 for categorical and numerical ex-situ representations, respectively).
Participants expressed more surprise about how their interactions and selections mapped to the underlying dataset when considering the \textit{summative} view, suggesting that the view increased their awareness of bias in their analysis process (e.g., P10-I said ``I'm surprised I didn't choose a doctor'').

\subsection{In-Lab Study 3}
In the third formative study, again 24 participants utilized the Control and Intervention interfaces to choose a political committee. 
This study focused on qualitative analysis of \emph{awareness}, while also addressing some shortcomings of the previous experiment (namely, the previous experiment was completed primarily by male participants, and the dataset used had only one female Republican).
\revised{We observed similar, yet weaker, effects as the previous in-lab study.}
\delete{While we find some weak evidence for the previously observed trends in behavior and decision making, in this experiment, the significance of quantitative findings did not replicate.}
It could be that there is \delete{simply} \revised{weak or} no effect (which is plausible given the exploratory nature of our analyses), or it could be the result of a confounding change to the interface in this study. 
In particular, for the study, we permitted categorical attributes to be assigned to axes of the scatterplot. 
The result is that well-formed clusters appear on the scatterplot, which could itself \delete{serve a bias-mitigating purpose} \revised{help people more easily choose representative samples} (e.g., pick a point from each cluster). 

We again observed qualitative evidence of the efficacy of interaction traces toward increasing awareness of potential biases. 
Because the interaction trace view compares a user's interactions to the underlying distribution of the data, we hypothesized this would lead to changes in user \delete{behavior} \revised{decision making} to \textit{make the committee more proportionally representative of the underlying dataset}. 
For example, one participant's committee was comprised of 10 Democrats, until interacting with the interaction trace view. 
The participant then adjusted the committee from 10 Democrats to 4 Republicans and 6 Democrats. 
In fact, examination of the interaction trace view made the participant aware of a mistake in her analysis: ``I forgot I had only filtered by Democrats.''

Also consistent with previous findings, \textbf{participants indicated higher preference for \emph{summative} interaction traces over \emph{real-time} interaction traces}. 
Further, using a grounded theory approach to code participant utterances during think-aloud \emph{summative} review of interaction traces, we found that \textbf{participants in the Control condition made more statements on average indicating heightened \emph{awareness} than participants in the Intervention condition}. 
We hypothesize this may be due to the fact that Intervention participants already saw their interaction traces in \textit{real-time} prior to the summative review phase.

\subsection{Summary} 
As a result of these observations, we planned a fourth experiment to be conducted virtually with crowd-sourcing, in which we made the following adjustments: 
\begin{enumerate}[nosep]
\item We adjusted the size and dimensionality of the dataset to be sufficiently difficult such that unconscious biases may arise, 
\item We corrected for confounds resulting from interface changes between the formative studies (in the crowd-sourced experiment, we permit numerical and ordinal attributes only to be assigned to axes, while categorical attributes can be used to apply filters; implications of which are described further in the Discussion).
\end{enumerate}

We also added an additional task in the domain of movies to study the generalizability of our approach.
Furthermore, across all of the in-lab formative studies, the vast majority of our participants identified as Democrats.
Hence, conducting an experiment via the online crowd-sourcing platform Amazon Mechanical Turk allowed us to broaden the political demographic of our users. 

\section{Crowd-Sourced Experiment}
\label{sec:results}


\subsection{Procedure}
\label{sec:procedure}
This study utilized a 2x2 design which manipulated real-time interaction traces (present, absent) x summative interaction traces (before revision, after \revised{revision}).
Participants in the user study were randomly assigned to one of four conditions: \ctrl, \summ, \rt, or \rtsum{} (Figure~\ref{fig:study_summary}I).
The procedure is depicted in Figure~\ref{fig:study_summary}.
After providing informed consent, participants completed a background questionnaire. 
Participants were shown a demonstration video of the interface using a cars dataset, then given the opportunity to practice by \textit{choosing a shortlist of 5 cars they would be interested to test drive}. 

Participants completed the first task (either politics or movies) followed by the second (movies or politics), with the order counterbalanced between subjects.
For each task, participants first chose a set of 10 politicians / movies, then submitted their decision.
Next, participants were either immediately given the opportunity to revise their selection (\ctrl{} and \rt) or were shown the summative interaction trace view (\summ{} and \rtsum).
The summative interaction trace view shown in Figure~\ref{fig:study_summary}(G-H) depicted for each attribute of the dataset: the underlying distribution (gray), and the distribution of user interactions (blue). 
Then, based on any imbalances observed, participants were given the opportunity to reflect and revise their committee if desired.
Lastly, those who did not see the summative interaction trace view before revision (\ctrl{} and \rt) were shown the view at the end after their decision was finalized. 

Those who saw summative interaction traces \emph{before} revision could incorporate any findings or realizations about their analysis process into subsequent revision, while those who saw summative interaction traces \emph{after} revision could only use this information to reflect afterwards, without impacting any decisions. 
The study took participants 44 minutes on average, and they were compensated \$10. 


\subsection{Participants} 
\label{sec:participants}
Based on a statistical power analysis from formative studies, we determined that at least 11 participants per condition would be required to detect an effect ($power = 0.8$, $\alpha = 0.05$). 
We recruited 56 participants via Amazon Mechanical Turk. 
We ultimately rejected 6 submissions due to a combination of failed attention checks, missing data, speeding through the study, and poor open-ended responses, leaving us with data from 50 participants who were randomly assigned to one of four conditions (13 \ctrl, 14 \summ, 11 \rt, 12 \rtsum).
Workers were restricted to only those located in the U.S. state of Georgia with 5,000+ approved HITs over their lifetime and a $\geq$ 97\% approval rating.
By gender, participants identified as female (28), male (21), and agender (1). 
Participants were 24-69 years old ($\mu = 40$, 1 preferred not to say) and self-reported an average visualization literacy of $\mu = 3.08, \sigma = 0.9$ on a 5-point Likert scale.
They had a wide range of educational backgrounds (6 high school, 12 some college, 7 associate's degree, 20 bachelor's degree, 3 master's degree, and 2 post-graduate degree); and a variety of fields of work, including e.g., art, bookkeeping, computer science, criminal justice, marketing, microbiology, office administration, political science, sales, social work, among others.
We refer to participants from each condition as \{$P_{CTRL}1$ - $P_{CTRL}13$\}, \{$P_{SUM}1$ - $P_{SUM}14$\}, \{$P_{RT}1$ - $P_{RT}11$\}, and \{$P_{RT+SUM}1$ - $P_{RT+SUM}12$\}.

\smallskip
\noindent\textbf{Politics Background. } 
Most participants had voted in US Presidential (49), state (43), or local (35) elections.
Participants identified as Democratic (35) and Republican (15), rating themselves on the political spectrum as conservative (2), moderate conservative (9), moderate (8), moderate liberal (14), and liberal (17). 

\smallskip
\noindent\textbf{Movies Background. }
Participants rated varying importance of movies in their lives (2 no importance, 15 little importance, 20 moderate importance, 12 large importance, 1 most importance). 
They watched movies daily (4), weekly (30), or monthly (16) and reported a diverse range of preferred genres.

\subsection{Hypotheses}
\label{sec:hypotheses}
Based on findings from formative in-lab studies, our hypotheses for this experiment are as follows. 
We organize our hypotheses according to those regarding \textbf{B}ehavior, \textbf{D}ecisions, \textbf{A}wareness, and \textbf{U}sability. 
\begin{itemize}[nosep]
    \item [\textbf{B1}] \emph{Real-time} interaction traces will have an effect on users' analysis \emph{process}. 
    \item [\textbf{B2}] \emph{Summative} interaction traces seen \emph{before} revision will lead users to make more \emph{revisions}. 
    \item [\textbf{D1}] \summ, \rt, and \rtsum{} participants will make selections more proportional to the underlying data than \ctrl{} participants. 
    \item [\textbf{A1}] \ctrl{} participants will exhibit greater \emph{surprise} upon seeing \emph{summative} interaction traces (Figure~\ref{fig:study_summary}(G-H)) than participants in \summ, \rt, and \rtsum.
    \item [\textbf{A2}] The attributes for which participants indicate \emph{surprise} about their interaction traces will correlate to lower \emph{focus}.
    \item [\textbf{A3}] The attributes for which participants indicate \emph{surprise} or \emph{focus} about their interaction traces will correlate to AD metric values.
    \item [\textbf{U1}] Participants in the \rt{} and \rtsum{} conditions will \emph{not} consistently use the real-time interaction trace view (Figure~\ref{fig:study_summary}F). 
    \item [\textbf{U2}] Participants will find the summative interaction trace visualization (Figure~\ref{fig:study_summary}(G-H)) more useful than real-time interaction trace visualizations (Figure~\ref{fig:study_summary}(E-F)). 
\end{itemize}

\vskip-\lastskip\delete{Our analyses are primarily quantitative.}
\revised{Based on guidance for statistical communication~\cite{dragicevic2016fair}, our analyses relied primarily on parameter estimation for all quantitative measures, using empirical bootstrapping with 1000 resamples to estimate the 95\% confidence intervals around all sample means}.
\delete{We used One-way ANOVA for all subsequent significance measures when comparing values across conditions and across levels of focus and surprise, described further in Section~\ref{sec:awareness}.}
We prioritize reporting results for attributes that users indicated as high focus (e.g., the top three are \attr{Ban Abortion After 6 Weeks}, \attr{Party}, and \attr{Gender} for the political task; \attr{IMDB Rating}, \attr{Genre}, and \attr{Rotten Tomatoes Rating} for the movies task\delete{; Figure~\ref{fig:focus_summary}}). Complete analyses are available in supplementary materials.


\subsection{Behavior}
We hypothesized that the presence of \emph{real-time} interaction traces would impact user behavior as measured by (1) interaction counts, (2) bias metric values~\cite{WallBias}, and (3) revisions during Phase 2. 

\begin{figure}[t]
  \centering
  \includegraphics[width=\columnwidth]{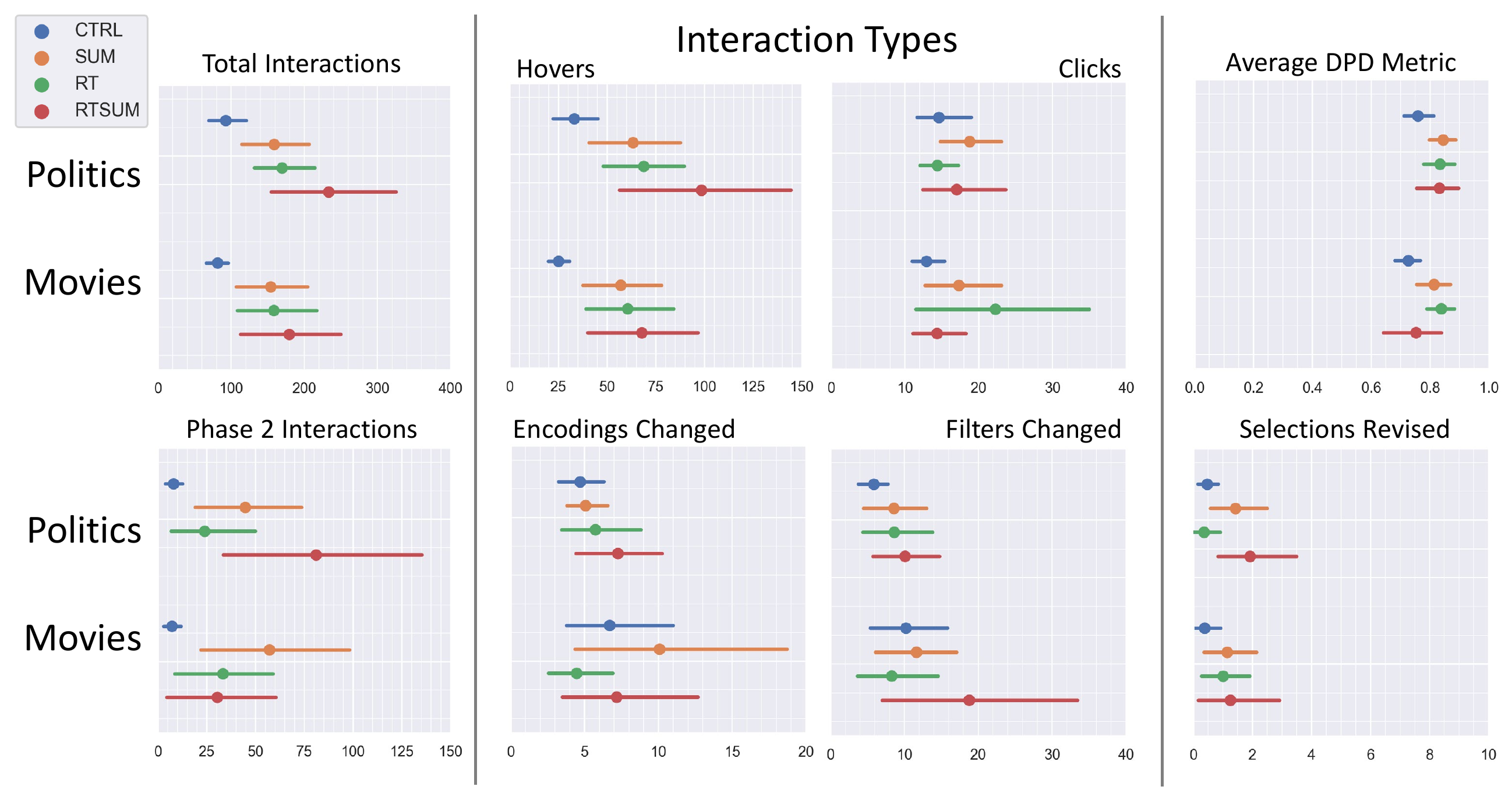}
  \caption{\revised{Modeling user behavior using point and interval estimation of the mean value of interactions of different types performed, revisions made, and the DPD bias metric.}}
  \label{fig:interactions_pointplots}
  \vspace{-1em}
\end{figure}

\smallskip
\noindent\textbf{Interactions. }
\revised{Figure~\ref{fig:interactions_pointplots} (left, center) illustrates total number and type of interactions performed by users in each condition for both the politics and movie tasks. 
We find some notable distinctions as observed by lesser overlap in confidence intervals.
Namely, \ctrl{} participants performed fewer hover interactions and fewer total interactions, demonstrating less interactive behavior than those who saw interaction traces. However, other specific interaction types showed less obvious trends.}
\delete{
We observed significantly fewer hover interactions from \ctrl{} participants in the movies task ($\mu_{CTRL} = 23.62$, $\mu_{SUM} = 55.29$, $\mu_{RT} = 56.73$, $\mu_{RT+SUM} = 66.50$; $\mathbf{p = 0.0324}$) and in the political task ($\mu_{CTRL} = 31.46$, $\mu_{SUM} = 60.36$, $\mu_{RT} = 67.18$, $\mu_{RT+SUM} = 96.17$; $\mathbf{p = 0.0201}$), indicating a lesser number of discrete exploratory interactions with individual data points.
We observed some significant difference in number of other types of interactions, including click interactions (i.e., adding or removing committee members), number of encodings changed (i.e., assigning attributes to axes), or filter interactions (i.e., adding, removing, or changing filter values) for the movies task: \textbf{encodings} ($\mu_{CTRL} = 0.74$, $\mu_{SUM} = 1.12$, $\mu_{RT} = 0.49$, $\mu_{RT+SUM} = 0.80$; $\mathbf{p = 0.0628}$) and \textbf{filters} ($\mu_{CTRL} = 1.14$, $\mu_{SUM} = 1.29$, $\mu_{RT} = 0.92$, $\mu_{RT+SUM} = 2.08$; $\mathbf{p = 0.0785}$), but not \textbf{clicks} ($\mu_{CTRL} = 11.62$, $\mu_{SUM} = 14.31$, $\mu_{RT} = 16.91$, $\mu_{RT+SUM} = 12.25$; $p = 0.2762$); and no significant difference for any interaction types for the politics task: \textbf{clicks} ($\mu_{CTRL} = 12.77$, $\mu_{SUM} = 15.21$, $\mu_{RT} = 12.82$, $\mu_{RT+SUM} = 13.67$; $p = 0.4794$), \textbf{encodings} ($\mu_{CTRL} = 0.52$, $\mu_{SUM} = 0.56$, $\mu_{RT} = 0.64$, $\mu_{RT+SUM} = 0.81$; $p = 0.1152$), or \textbf{filters} ($\mu_{CTRL} = 0.65$, $\mu_{SUM} = 0.95$, $\mu_{RT} = 0.96$, $\mu_{RT+SUM} = 1.12$; $p = 0.2334$).
There may be task-related effects that influence users' interactive analyses.} 
This result provides \textbf{some support for hypothesis B1}.

\begin{figure}[t]
  \centering
  \includegraphics[width=\columnwidth]{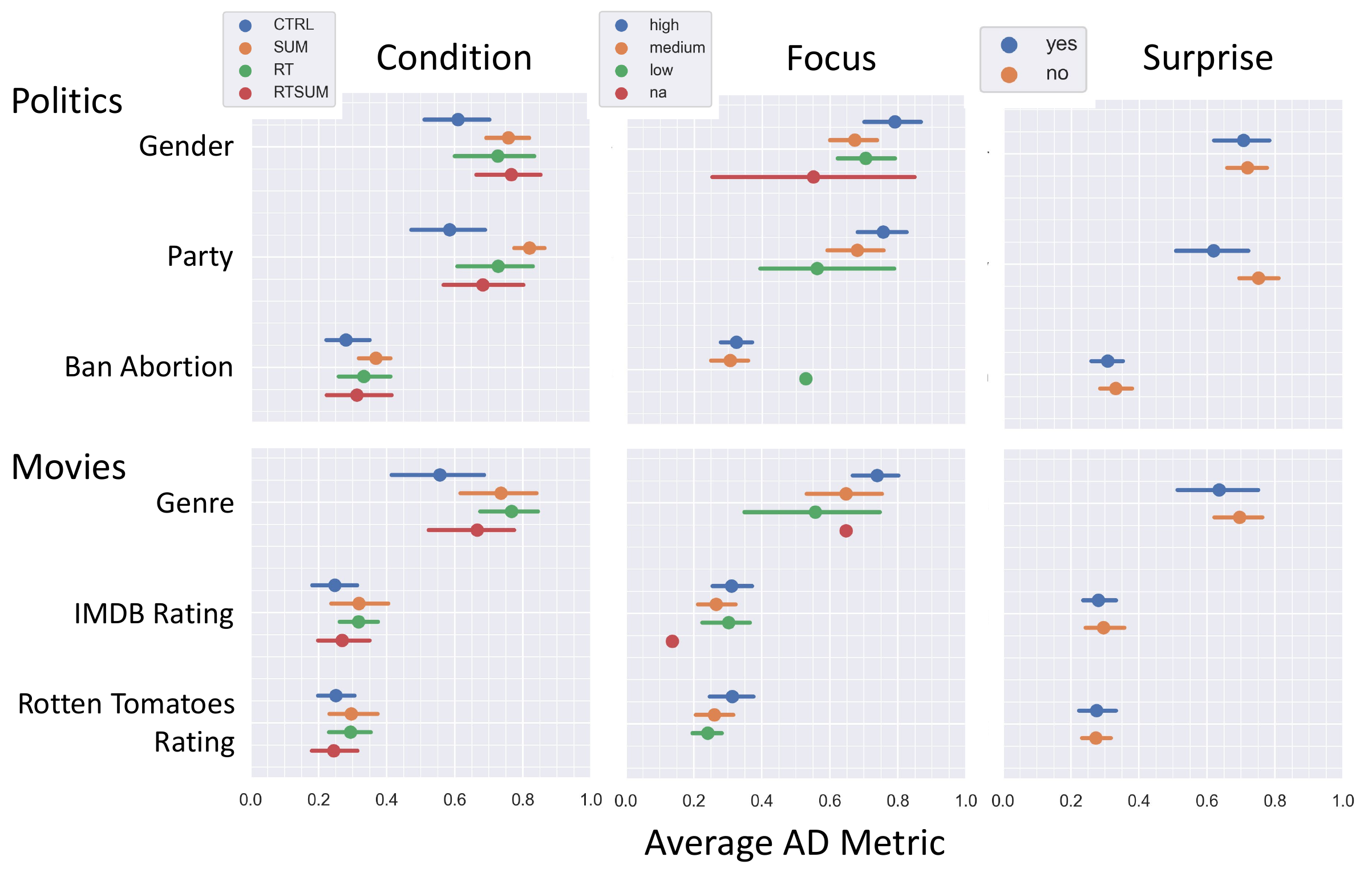}
  \caption{\revised{The effect of condition, focus, and surprise on estimating the average AD metric value for select attributes in each task.}}
  \label{fig:ad_metric_pointplots}
  \vspace{-1em}
\end{figure}

\smallskip
\noindent\textbf{Bias Metrics. }
The AD bias metric values provide one way of quantifying how a user's interactive behavior aligns with the distributions of the underlying data per attribute~\cite{WallBias}.
Lower metric values indicate interaction distributions that are more similar to the distribution of a given attribute, while higher metric values indicate dissimilarity. 
\revised{Figure~\ref{fig:ad_metric_pointplots} (left) shows the average AD bias metric values for the top three attributes that participants focused on in each task.
We observe that \ctrl{} participants exhibited less bias towards some attributes (e.g., \attr{Gender} and \attr{Party} in the political task and \attr{Genre} in the movies task) while other attributes display less clear trends.}
\delete{
We observe a difference between conditions for some attributes. 
For instance, Figure~\ref{fig:bias_ad_politics_movies}a shows the average AD bias metric values for the top three focused attributes in the political task. 
We find that \textbf{\attr{Party}} ($\mu_{CTRL} = 0.59$, $\mu_{SUM} = 0.82$, $\mu_{RT} = 0.73$, $\mu_{RT+SUM} = 0.68$; $\mathbf{p = 0.0207}$) and \textbf{\attr{Gender}} ($\mu_{CTRL} = 0.61$, $\mu_{SUM} = 0.76$, $\mu_{RT} = 0.73$, $\mu_{RT+SUM} = 0.77$; $\mathbf{p = 0.0864}$) indicate greater differences, while \textbf{\attr{Ban Abortion After 6 Weeks}} has no substantial difference ($\mu_{CTRL} = 0.28$, $\mu_{SUM} = 0.37$, $\mu_{RT} = 0.33$, $\mu_{RT+SUM} = 0.31$; $p = 0.3840$).
Similarly, for the movies task (Figure~\ref{fig:bias_ad_politics_movies}b), we find differences for \textbf{\attr{Genre}} ($\mu_{CTRL} = 0.56$, $\mu_{SUM} = 0.74$, $\mu_{RT} = 0.77$, $\mu_{RT+SUM} = 0.67$; $\mathbf{p = 0.0971}$), but not for \textbf{\attr{IMDB Rating}} ($\mu_{CTRL} = 0.25$, $\mu_{SUM} = 0.32$, $\mu_{RT} = 0.32$, $\mu_{RT+SUM} = 0.27$; $p = 0.4913$) or \textbf{\attr{Rotten Tomatoes Rating}} ($\mu_{CTRL} = 0.25$, $\mu_{SUM} = 0.30$, $\mu_{RT} = 0.29$, $\mu_{RT+SUM} = 0.24$; $p = 0.6077$).
}
\delete{Notably, \ctrl{} participants actually had \emph{lower} AD bias metric values than participants in other conditions.}
We hypothesize that, with increased \emph{awareness of unconscious biases}, the interaction trace interventions may have ultimately \emph{amplified conscious biases}, \revised{discussed further in Section~\ref{sec:discussion}}. 

The DPD bias metric similarly quantifies how evenly a user's interactions are divided among individual data points.
\revised{Figure~\ref{fig:interactions_pointplots} (top, right) shows a slight trend toward lower DPD metric values for \ctrl{} participants.}
\delete{
We again do not find a significant difference between conditions for either the political task ($\mu_{CTRL} = 0.76$, $\mu_{SUM} = 0.84$, $\mu_{RT} = 0.83$, $\mu_{RT+SUM} = 0.83$; $p = 0.1512$) or movies task ($\mu_{CTRL} = 0.73$, $\mu_{SUM} = 0.81$, $\mu_{RT} = 0.84$, $\mu_{RT+SUM} = 0.75$; $p = 0.1081$).
}
These results provide \revised{\textbf{mixed support for hypothesis B1}.} \delete{\textbf{some support for hypothesis B1}}

\smallskip
\noindent\textbf{Revisions. }
All participants had the opportunity to revise their initial selections. 
We hypothesized that those participants who saw \emph{summative} interaction traces \emph{before} revision would make more revisions (i.e., number of edits to their initial selection after Phase 1). 
\revised{Figure~\ref{fig:interactions_pointplots} (bottom right) shows that \summ{} and \rtsum{} participants who saw summative interaction traces before revision tended to make more revisions compared to \ctrl{} and \rt{} participants during the political task, but not during the movies task.
Similarly, the same groups tended to perform more interactions in general in the phase 2 revision (Figure~\ref{fig:interactions_pointplots} (bottom left)), demonstrating that summative interaction traces tended to correspond to more exploration and decision changes during revision.}
\delete{
We find support for this hypothesis for the politics task ($\mu_{CTRL} = 0.46$, $\mu_{SUM} = 1.43$, $\mu_{RT} = 0.36$, $\mu_{RT+SUM} = 1.92$; $\mathbf{p = 0.0699}$), and while the trend is generally consistent for the movies task, it is not significant ($\mu_{CTRL} = 0.38$, $\mu_{SUM} = 1.14$, $\mu_{RT} = 1.00$, $\mu_{RT+SUM} = 1.25$; $p = 0.6318$). 
In addition to making more revisions to their selections, participants who saw summative interaction traces before revision also explored the data more (i.e., performed more interactions) before submitting their final selection with a stronger effect in the politics task ($\mu_{CTRL} = 8$, \textbf{$\mu_{SUM} = 45$}, $\mu_{RT} = 24$, \textbf{$\mu_{RT+SUM} = 81$}; $\mathbf{p = 0.0156}$) compared to the movies task ($\mu_{CTRL} = 7$, \textbf{$\mu_{SUM} = 57$}, $\mu_{RT} = 33$, \textbf{$\mu_{RT+SUM} = 30$}; $p = 0.1199$).}
This result \textbf{confirms hypothesis B2}.


\subsection{Decisions}
We hypothesized that participants in \summ, \rt, and \rtsum{} (i.e., those who were influenced by interaction traces in some format, real-time or summative, before finalizing their decisions) would ultimately make choices that were more proportional to the underlying data compared to participants who did not (\ctrl). 
We quantify this effect for binary attributes in the datasets (e.g., \attr{Party} and \attr{Gender} in the politics dataset), by considering the ratio of values \emph{chosen} with respect to the ratio that appears in the \emph{dataset}. 
In the movies task, no attributes are binary; hence this analysis only applies to \attr{Party} and \attr{Gender} in the politics task.

\begin{figure}[t]
   \centering
   \includegraphics[width=\columnwidth]{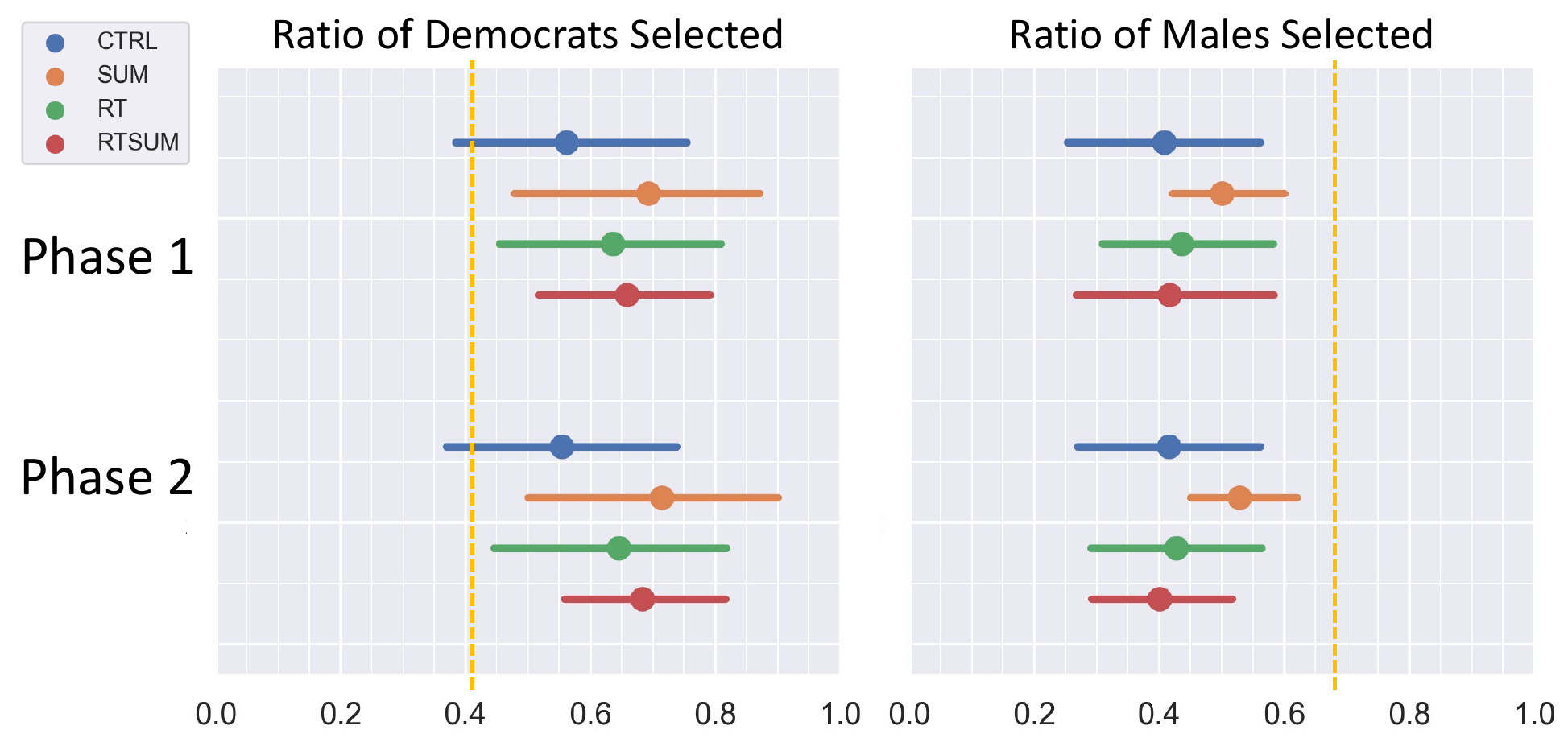}
   \caption{\revised{Mean estimates for the ratio of Democrats (left) and Males (right) in participants' chosen committees, selected during each phase for each condition. Vertical dashed lines indicate ratio of each candidate type present in the underlying dataset.}}
   \label{fig:ratios_pointplots}
   \vspace{-1em}
\end{figure}

For \attr{Party}, the dataset contains 59\% Republicans and 41\% Democrats.
For \attr{Gender}, the dataset contains 32\% females and 68\% males.
\revised{As shown in Figure~\ref{fig:ratios_pointplots}, \emph{all conditions} chose committees that were relatively dissimilar from the underlying distributions of \attr{Gender} and \attr{Party} in the dataset (annotated with a vertical dashed line).
We discuss this result further in Section~\ref{sec:discussion}.
In fact, for participants who saw any intervention (\rt, \summ, \rtsum), they trended toward choosing \emph{more dissimilar} distributions of \attr{Party} compared to \ctrl{} participants.
However, there are no clear distinctions in the ratios of \attr{Gender} or \attr{Party} between conditions.
Hence, we find \textbf{no support for hypothesis D1}.}
\delete{We found that \ctrl{} participants chose, on average, 48\% males and 55\% Democrats in their committees, whereas \summ, \rt, and \rtsum{} collectively chose, on average 57\% males and 68\% Democrats in their committees.
\ctrl{} participants were \emph{slightly} closer to the expected number of proportional choices for \attr{Gender} but more skewed on average for \attr{Party}.
However, we find that \emph{all} conditions made choices of significantly different \attr{Gender} and \attr{Party} composition than the underlying data.
Using a One-Sample t-Test, we compared the distribution of ratios across each of \ctrl{} and \summ, \rt, and \rtsum{} collectively to the baseline data representation above.
The deviation between expectation and observation was statistically significant across the board: both Male selection ratio in \ctrl{} ($t = -3.1418$, $\mathbf{p = 0.0043}$) and \summ, \rt, and \rtsum{} ($t = -3.2300$, $\mathbf{p = 0.0019}$) collectively; as well as Democrat selection ratio in \ctrl{} ($t = 1.5337$, $p = 0.1510$) and \summ, \rt, and \rtsum{} ($t = 5.2157$, $\mathbf{p = 0.000008}$) collectively.
Comparing between all conditions (\ctrl, \summ, \rt, \rtsum) using ANOVA, we found no statistical significance in the difference between Male ratio ($p = 0.4141$) or Democrat ratio ($p = 0.6272$).
This result provides \textbf{no support for hypothesis D1}.
}

\begin{figure}[t]
   \centering
   \includegraphics[width=\columnwidth]{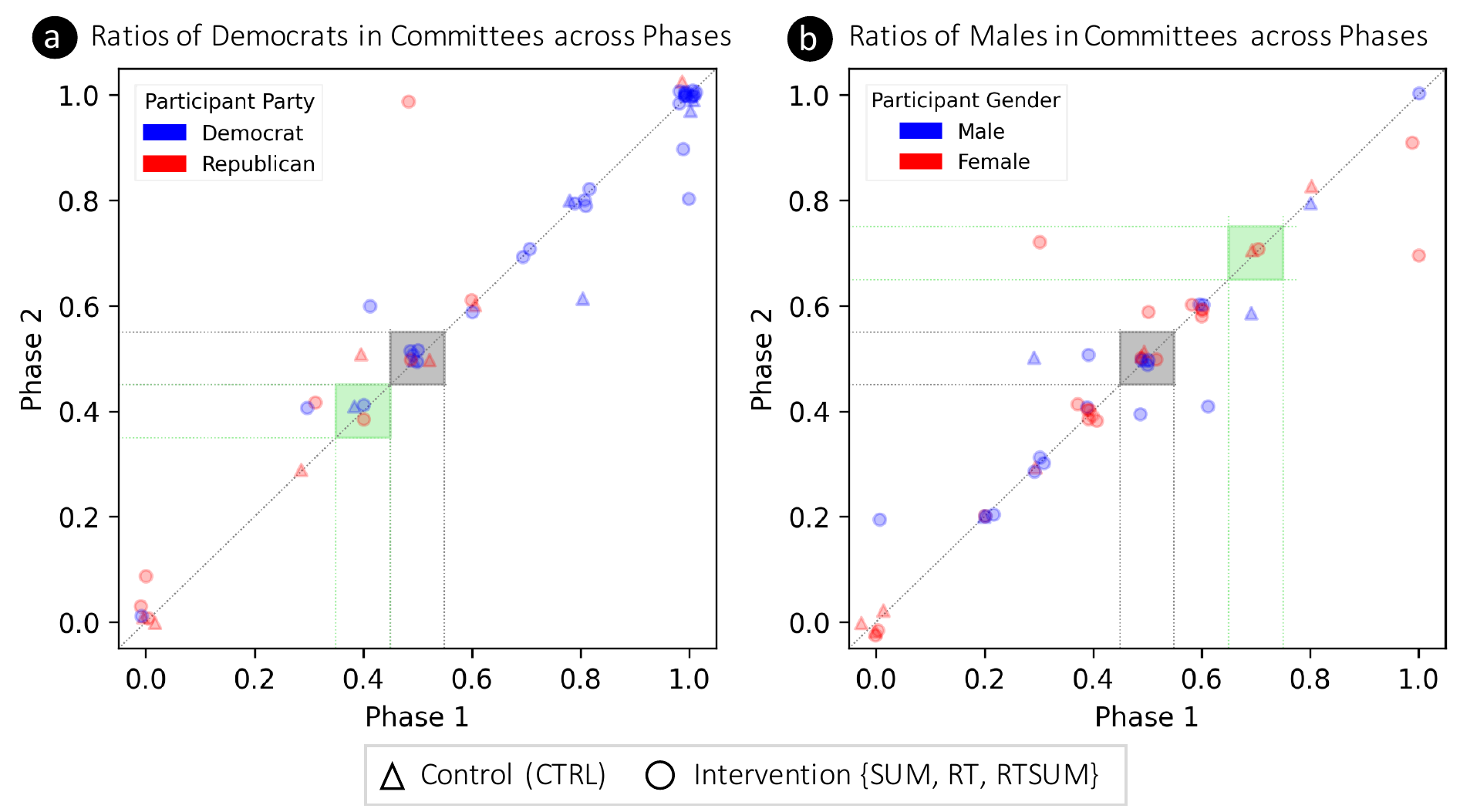}
   \caption{Ratio of \attr{Party} (a) and \attr{Gender} (b) composition of committees in Phase 1 and Phase 2 of the political task, colored by the participant's corresponding party affiliation and gender.}
   \label{fig:ratio_party_gender}
   \vspace{-1em}
\end{figure}

While we observe no clear effects on committee composition based on intervention conditions, we do see some distinctions in the way participants chose their committee based on their own political party affiliation. 
For instance, Figure~\ref{fig:ratio_party_gender}a shows the ratio of Democrats that participants chose in Phase 1 (x-axis) and in Phase 2 (y-axis) committees. 
Points that fall on the diagonal represent participants who did not change the composition of their committee by \attr{Party} during the revision. Points are colored by the political party affiliation of the participant.
There is clear delineation, where participants who most identified with Democrats (blue) appear in the top right (choosing Democrat-dominant committees) and participants who most identified with Republicans (red) appear in the bottom left (choosing Republican-dominant committees). 
Three notable examples emerge e.g., a blue circle in the bottom left, a red circle in the top middle, and a red triangle in the top right. 
These participants chose final committees that were entirely composed of politicians from the opposite \attr{Party} than their own affiliation. 
Upon further inspection, two participants identified as moderate or neutral on the spectrum of liberal $\rightarrow$ conservative, and one expressed divergent beliefs from their affiliated party. 
For \attr{Gender} (Figure~\ref{fig:ratio_party_gender}b), we do not observe such a clear distinction. 
Participants' committees reveal some clusters (e.g., in the bottom left, many female participants chose committees of all female politicians; just above that cluster along the diagonal, many men chose committees with 70-80\% female politicians; etc). 
However, there is less strict division in the overall trend ($\mu_{Female} = 0.45$, $\mu_{Male} = 0.44$).


\subsection{Awareness}
\label{sec:awareness}
While formative studies indicated variable impacts of interaction traces on behavior and decisions, they had a more promising qualitative effect on people's \emph{awareness} of potential unconscious biases that drive data analysis and decision making. 
In particular, we assess awareness by asking two questions for each attribute of the data, at the time that the user viewed the summative interaction traces (\emph{before revision} for \summ{} and \rtsum, and \emph{after final selections} for \ctrl{} and \rt).
\begin{enumerate}[nosep]
    \item Are you \textbf{surprised} by your interactions with this attribute? \{\emph{yes}, \emph{no}\}
    \item How much \textbf{focus} did you give this attribute during your task? \{\emph{high}, \emph{medium}, \emph{low}, \emph{NA}\}
\end{enumerate}

\begin{figure}[t]
  \centering
  \includegraphics[width=\columnwidth]{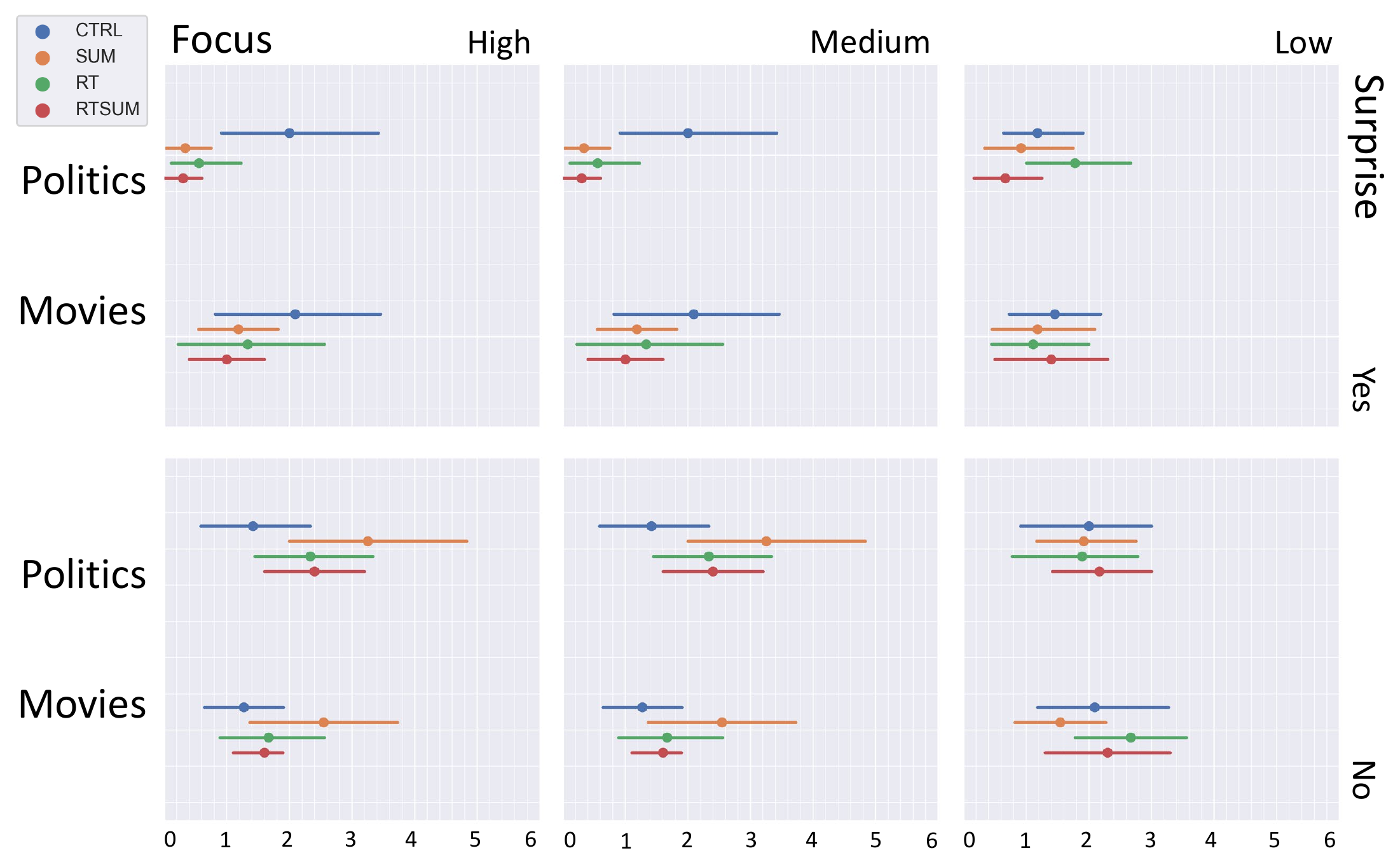}
  \caption{\revised{Estimating the mean number of times a combination of focus and surprise was recorded for each task and condition.}}
  \label{fig:surprise_focus_pointplots}
  \vspace{-1em}
\end{figure}

\revised{Figure~\ref{fig:surprise_focus_pointplots} compares the average number of times a particular combination of focus and surprise was recorded between all conditions and tasks.} 
We hypothesized that \ctrl{} participants would express \delete{greater} surprise \revised{more often} upon seeing the \emph{summative} interaction traces than participants in the other conditions. 
Participants in all other conditions had some form of signal from interaction traces (real-time or summative) \emph{before} their final decisions, whereas \ctrl{} participants only saw the summative view after their final selections were locked in.
\delete{
As Figure~\ref{fig:surprise_focus} demonstrates by the higher red bars for (a), \ctrl{} participants in the movies task expressed surprise towards an attribute 50\% of the time, compared to other conditions: \summ{} (b) at 33\%, \rt{} (c) at 33\%, and \rtsum{} (d) at 38\%.}
\revised{
In the political task, we observe that \ctrl{} participants expressed surprise for high-focus attributes more often than the other conditions, and conversely reported the lowest numbers of no surprise for high-focus attributes (Figure~\ref{fig:surprise_focus_pointplots} left). 
This provides some support for the idea that seeing summaries, either in real-time (\rt{}), before revisions (\summ{}) or both (\rtsum{}), reduces the surprise reported when participants expressed high focus in more divisive analysis scenarios.
The effect was less pronounced in the movies task.
In general, as focus decreased on any given attribute, the number of surprised responses decreased uniformly as well across conditions and tasks.
This pattern further supports the idea that summaries can affect whether participants are aware of their analysis strategies at the end of a task.}
This result \textbf{confirms hypothesis A1}. 

We also hypothesized that \delete{the attributes participants explicitly focused on} \revised{expressing lower focus on an attribute} (e.g., \attr{Ban Abortion After 6 Weeks}, \attr{Party}, and \attr{Gender} for the political task; \attr{IMDB Rating}, \attr{Genre}, and \attr{Rotten Tomatoes Rating} for the movies task\delete{, Figure~\ref{fig:focus_summary}}) would be correlated to \delete{greater} \revised{more instances of} surprise. 
That is, we believed attributes that were unattended may have a surprising distribution of user interactions, since the user did not focus on them. 
\delete{
However, we find no trend in the political task (Figure~\ref{fig:surprise_focus_total}a) or the movies task (Figure~\ref{fig:surprise_focus_total}b).}
\revised{Comparing each column of Figure~\ref{fig:surprise_focus_pointplots} top v. bottom demonstrates that when \ctrl{} participants expressed surprise (top, blue), fewer people expressed that level of focus compared to when participants were not surprised (bottom, blue).
For other conditions, there does not appear to be any substantial difference in surprise (top) v. no surprise (bottom) w.r.t. focus. 
That is, surprise and focus do not appear to be correlated. 
All means and intervals are roughly between 1 and 3.}
This result \textbf{disconfirms hypothesis A2}.


Lastly, we hypothesized that there would be a correlation between attributes that participants' \emph{focused} on \revised{(Figure~\ref{fig:ad_metric_pointplots} center)} or were \emph{surprised} by \revised{(Figure~\ref{fig:ad_metric_pointplots} right)} and the average AD bias metric values.
\revised{Some attributes (e.g., \attr{Party} in the political task, \attr{Rotten Tomatoes Rating} in the movies task) roughly corresponded to greater focus related to higher AD bias metric values.
On the other hand, some attributes for which participants expressed surprise about their interactions (e.g., \attr{Party} in the political task) corresponded to lower AD bias metric values 
However, these trends were not true for all attributes, hence \textbf{support for hypothesis A3 is inconclusive}.}
\delete{
In the political task, participants who expressed high / medium / low focus on some attributes had correspondingly high / medium / low AD bias metric values on average for \attr{Gender} ($\mu_{high} = 0.79$, $\mu_{medium} = 0.67$, $\mu_{low} = 0.70$; $\mathbf{p = 0.0875}$), but not for \attr{Party} ($\mu_{high} = 0.76$, $\mu_{medium} = 0.68$, $\mu_{low} = 0.56$; $p = 0.1115$) or \attr{Ban Abortion After 6 Weeks} ($\mu_{high} = 0.32$, $\mu_{medium} = 0.32$, $\mu_{low} = 0.53$; $p = 0.2846$) in the political task (Figure~\ref{fig:bias_ad_surprise_focus}c). 
Similarly, participants who expressed yes / no surprise on some attributes had corresponding high / low AD bias metric values on average (i.e., they were surprised to see the corresponding bias value in the summary view, Figure~\ref{fig:bias_ad_surprise_focus}a) for \attr{Party} ($\mu_{yes} = 0.62$, $\mu_{no} = 0.75$; $\mathbf{p = 0.0282}$) but not for \attr{Gender} ($\mu_{yes} = 0.71$, $\mu_{no} = 0.72$; $p = 0.8254$) or \attr{Ban Abortion After 6 Weeks} ($\mu_{yes} = 0.31$, $\mu_{no} = 0.33$; $p = 0.5736$).
}

\delete{
In the movies task, on the other hand, we observed no trend toward participants who expressed high / medium / low focus with corresponding high / medium / low AD bias metric values on average (Figure~\ref{fig:bias_ad_surprise_focus}d) for \attr{IMDB Rating} ($\mu_{high} = 0.31$, $\mu_{medium} = 0.27$, $\mu_{low} = 0.30$; $p = 0.5976$), \attr{Genre} ($\mu_{high} = 0.74$, $\mu_{medium} = 0.65$, $\mu_{low} = 0.56$; $p = 0.1151$), and \attr{Rotten Tomatoes Rating} ($\mu_{high} = 0.31$, $\mu_{medium} = 0.26$, $\mu_{low} = 0.24$; $p = 0.2179$).
Similarly, none of the participants who expressed yes / no surprise on some attributes had corresponding high / low AD bias metric values on average (i.e., they were surprised to see the corresponding bias value in the summary view, Figure~\ref{fig:bias_ad_surprise_focus}b) for \attr{IMDB Rating} ($\mu_{yes} = 0.28$, $\mu_{no} = 0.30$; $p = 0.7151$), \attr{Genre} ($\mu_{yes} = 0.64$, $\mu_{no} = 0.70$; $p = 0.3921$), and \attr{Rotten Tomatoes Rating} ($\mu_{yes} = 0.28$, $\mu_{no} = 0.27$; $p = 0.9511$).
Hence, based on minimal support in the political task and no support in the movies task, we find \textbf{support for hypothesis A3 is inconclusive}.
}

\subsection{Usability}
Based on formative studies, we formulated two general hypotheses about the usability of interaction trace interventions. 
First, in formative studies we observed relatively little use of the \emph{real-time} interaction traces in the form of interactions with the Distribution panel (Figure~\ref{fig:study_summary}F), which we believe to be due to high cognitive load during the task itself (i.e., participants were unable to attend to an additional view in the system while trying to explore the data).
Indeed, participants interacted minimally\delete{, on average less than once,} with the Distribution panel in both the politics \delete{($\mu_{RT} = 0.23$, $\sigma_{RT} = 0.53$, $\mu_{RT+SUM} = 0.15$, $\sigma_{RT+SUM} = 0.49$; $p = 0.2368$)} \revised{($\mu_{RT} = 2.09$, 95\% CI $[0.64, 3.82]$, $\mu_{RT+SUM} = 1.33$, 95\% CI $[0.33, 2.42]$)} and movies tasks \delete{($\mu_{RT} = 0.35$, $\sigma_{RT} = 0.73$, $\mu_{RT+SUM} = 0.23$, $\sigma_{RT+SUM} = 0.59$; $p = 0.1866$).} \revised{($\mu_{RT} = 3.18$, 95\% CI $[0.45, 6.28]$, $\mu_{RT+SUM} = 2.08$, 95\% CI $[0.17, 4.67]$).} 
This result \textbf{confirms hypothesis U1}.

For similar reasons (high cognitive load), we hypothesized that participants would prefer the \emph{summative} interaction traces over the \emph{real-time} interaction traces. 
This result was somewhat variable. 
\ctrl{} participants saw only summative interaction traces (Figure~\ref{fig:study_summary}G-H, after revision) and rated their utility on a Likert scale a median 4 / 5 across both politics and movies tasks. 
\summ{} participants who saw only the summative interaction traces (before revision) likewise gave a median 4 / 5 Likert rating for movies and 3.5 / 5 for politics.
\rt{} participants rated in-situ interaction traces (Figure~\ref{fig:study_summary}E), ex-situ interaction traces (Figure~\ref{fig:study_summary}F), and summative interaction traces (after revision) all the same with a median Likert rating of 4 / 5 across both politics and movies tasks. 
These participants all viewed \emph{summative} interaction traces very positively.
Only \rtsum{} participants differed, rating in-situ interaction traces 4 / 5 (politics and movies), ex-situ interaction traces 2 / 5 (politics) and 3 / 5 (movies), and summative (before revision) 3 / 5 for both politics and movies. 
The participants in the only condition that could compare both real-time and summative interaction traces surprisingly preferred \emph{real-time} interaction traces. 
Overall, \rtsum{} participants rated all forms of interaction trace lower than participants in the other conditions, perhaps due to this condition having the highest cognitive load of all the conditions, showing both real-time and summative interaction traces. 
This result \textbf{disconfirms hypothesis U2}.


\subsection{Qualitative Feedback}
The survey after each task included open-ended questions about the participant's decision criteria and, in the political scenario, their desired outcome of the committee. 
Below we discuss themes that emerged.

\smallskip
\noindent\textbf{Politics. }
Some focused on choosing diverse committees (e.g., $P_{CTRL}2$ said ``[they] tried to choose politicians with a wide range of views, to represent most people's views and not necessarily one side'').
Many expressed firm goals about choosing politicians differently than the underlying data distribution (e.g., $P_{CTRL}1$ expressed they wanted their committee to be mostly women because ``[they] truly feel abortion is an issue only women can really comment on''), while also simultaneously balancing other attributes (e.g., ``[they] also tried to get a mix of Republicans and Democrats with many years experience in politics to get a fair showing to both sides'').
Others sought committee members based on ``who [they] thought might share [their] values'' ($P_{CTRL}10$), instances of which were observed from both sides of the political spectrum (e.g., $P_{SUM}8$ chose ``[Republican] politicians who were in favor of the abortion ban'', while $P_{SUM}13$ intentionally ``chose all Democrats''). 

Ultimately, people hoped their chosen committee would lead to outcomes such as ``uphold the ban on abortion after 6 weeks, except in extreme cases where there is a life-threatening decision that needs to be made'' ($P_{RT}6$), that ``the abortion ban would become or stay law'' ($P_{SUM}12$), or ``assess the public opinion in an unbiased way'' ($P_{RT+SUM}2$). 
Many of the biases that emerged as a result of these goals were very conscious. 
For instance, $P_{RT}3$ expressed, ``I was really biased, to be honest. I wanted people who were in favor of abortion, and most, if not all, Republicans did not fit the bill.''
In some cases, participants adjusted their strategy for Phase 2 (Revision); e.g., $P_{RT+SUM}6$ said, ``my revised committee was solely focused on trying to get vast ideological perspective without necessarily focusing on men or women. I did more of a `blind' choosing without focusing so much on gender the second time around ... I only wanted different viewpoints without focusing too much on the extreme of either side.''


\smallskip
\noindent\textbf{Movies. }
Participants expressed diverse criteria for selecting movies. 
Some focused on making choices that ``were a good representation of the given dataset as a whole'' by ``find[ing] and select[ing] films that were spread out across the graph'' ($P_{CTRL}7$).
Several other participants expressed a focus on finding variety for one factor, e.g., highly rated ($P_{CTRL}6$) or diverse genres ($P_{CTRL}5$).
Others had criteria that were too abstractly expressed to capture in the data (e.g., ``selecting movies that [they] feel would be interesting to watch'' -$P_{CTRL}11$). 
Some participants expressed a focus on the dummy \attr{Title} attribute ($P_{RT}7)$.

Some participants expressed how interaction traces influenced their behavior. 
For instance, in Phase 2 (Revision), $P_{SUM}6$ expressed that ``the 2nd time through, [they were] careful to check the different areas such as genre'' to ``[assess] which [they] thought were representative in each case.''
Similarly, $P_{SUM}13$ was ``glad [they] got to go back and revise because [they] missed the creative type selection which [they] corrected to contemporary fiction.''
Others relied on different features to facilitate their analysis, e.g., filters ($P_{RT}3$).

\section{Discussion}
\label{sec:discussion}

\smallskip
\noindent\revised{\textbf{Conscious v. Unconscious Biases. }}
We observed both \delete{explicit} \revised{conscious} and \delete{implicit} \revised{unconscious} biases throughout the studies in this paper. 
\revised{While some results were as we hypothesized, there were a number of surprising findings (e.g., \ctrl{} participants exhibited lower DPD metric values and AD metric values for several attributes; \ctrl{} participants tended to pick more proportional political committees w.r.t. gender and party; surprise at seeing distributions in interaction traces corresponded to higher AD metric values; etc).
We speculate that some of these findings may be the result of interaction traces leading to \textbf{amplified conscious biases}. 
That is, while awareness of unconscious biases may have been improved, the same intervention may have led to exaggeration of conscious biases.
Additional studies are needed to understand the nature of the relationship.}
\delete{We hypothesized that participants would choose political committees and representative movies based on a few explicit criteria, and if dimensionality of data is high, that may result in \delete{implicit} \revised{unconscious} bias when people lose sight of other attributes.}
\revised{Regardless, t}hese \delete{implicit} \revised{unconscious} biases may be the result of lack of attention and unknown correlations in the data, or they could be the result of more dangerous implicit attitudes and stereotypes\delete{ that drive decision making, especially in the case of the political decision making scenario}. 
From a behavioral perspective, the interactions users perform related to \delete{explicit} \revised{conscious} or \delete{implicit} \revised{unconscious} bias may look similar. 
Thus in future work when in-lab experiments are again feasible (outside of the COVID-19 pandemic), eliciting user feedback can be helpful to refine models of bias by users directly indicating if their focus was intentional or not~\cite{WallDesignSpace} and by correlating outcomes with results of implicit association tests~\cite{greenwald1998measuring}.

\smallskip
\noindent\textbf{False Positives v. False Negatives. }
Given the imperfection of quantifying bias from user interactions, it begs the question: what is the harm of inaccuracy? 
A false positive (i.e., the system believes you are biased when you are not) could be frustrating to users, but we posit is relatively harmless apart from possible damaged ego. 
On the other hand, a false negative (i.e., the system believes you are not biased when you are) could be much more harmful, leading to unchecked errors. 
Furthermore, ``false negative'' circumstances are essentially the present norm, given that most systems do not attempt to capture bias in the analysis process.
In such circumstances, biases would have propagated unchecked regardless.
Hence we argue that a system that characterizes bias, even with low or unknown accuracy, can provide benefit in situations where bias may cause urgent problems.

\smallskip
\noindent\textbf{Implications of \delete{Balance} \revised{Bias} Definition. } 
The bias metrics~\cite{WallBias} used in these studies are formulated based on \revised{comparing user interactions to a \textbf{proportional} baseline.}
Given that visualizing these metrics in \textit{real-time} sometimes resulted in \delete{significant} changes in behavior and decisions, it begs the question whether this was the \emph{right} way to nudge participants.
Some participants expressed explicit goals to choose representative (or proportional) samples of political committees or movies, some aimed to choose equal samples (e.g., one movie from each \attr{Genre} or equal numbers of Democrats and Republicans), while still others intentionally biased their selections in other ways (e.g., all female politicians). 
Future work can explore the contexts in which different \revised{baselines of comparison} are appropriate \revised{given Wall et al.'s metrics~\cite{WallBias} or by introducing alternative metrics (e.g., that incorporate a user's prior beliefs using a Bayesian model).} \delete{, how visualizations can allow users to specify their targets, and how interfaces can be designed to nudge or encourage users to meet their target. }
This has further ethical implications, in that designers must take on the social responsibility to choose visualization designs and bias computation mechanisms that reflect social values without unduly compromising user agency. 

\smallskip
\noindent\textbf{Design Implications. }
Based on feedback in formative studies, with or without \textit{real-time} interaction trace visualizations, some participants found indirect ways to assess balance or bias in their committee choices (e.g., by applying filters and cycling through combinations of scatterplot axes to see the distribution of selected points). 
Hence, the affordances within the interface design can itself serve as a potentially powerful bias mitigation approach, promoting user awareness and enabling self-editing. 
Another example is enabling categorical attributes to be assigned to axes to see categorical distributions of selected points, which can offload oft complex management of cognitive decision making to a perceptual task. 
For instance, we varied this feature across formative studies and observed that categorical assignments enabled participants to easily identify clusters of points where they may not have selected any data points. 
Participants were able to choose \emph{equally} across clusters or intentionally \emph{bias} across clusters by visually inspecting for selected points.
Particularly in situations where cognitive overload may prevent users from managing secondary views, designing the interface to afford indirect assessment of their choices may be a better alternative.

\smallskip
\noindent\textbf{Study Limitations. }
Our formative studies suffered from a biased sampling of participants (mostly male Democrats). 
In the third and final formative study, we were able to correct for gender bias; however, due to sampling within our university, we were unable to recruit participants with diverse political party affiliations. 
We addressed these concerns in our fourth study using the crowdsourcing platform, Amazon Mechanical Turk. 
\revised{In spite of our best efforts, there was still a political imbalance (leaning Democrat), which we could not selectively recruit for due to our constraints of recruitment (within Georgia, high MTurk approval ratings).}
However, this experiment came with its own set of tradeoffs. 
While we were generally satisfied with participant engagement in the task as observed by their open-ended feedback, we were nonetheless limited in our ability to make rich observations. 
\revised{In addition, the task phrasing was intentionally vague to not bias participants toward any particular selection strategy. 
The cost, however, is noise in our data due to variable interpretations of the task. 
Future experiments may explore refining the task phrasing or exploring performance-based incentives to reduce the noise in collected user data.}
\section{Conclusion}
In this paper we explored the effect of \textit{real-time} and \textit{summative} visualization of user interaction traces toward mitigating human biases in decision making tasks in the domains of politics and movies, where success was measured by changes in (1) behavior, (2) decisions, and (3) awareness.
To study this effect, we conducted three formative in-lab experiments and a virtual crowdsourced experiment. 
We found that when both interventions were combined (real-time and summative), participants \revised{tended to perceive} \emph{both} to be less useful.
Hence, the impact of heightened awareness \revised{may} come at the expense of user experience. 
\revised{Furthermore, while we find some support for the impact of interaction traces (e.g., towards behavioral changes in interaction, for increasing awareness), we also find some surprising trends (e.g., \ctrl{} participants' lower bias metric values, slightly more skewed political committees). 
These mixed results suggest that while interaction traces may lead to increased awareness of unconscious biases, they may also lead to amplification of conscious biases.} 
\delete{We find that interaction traces 
influence users' behavior and increase awareness of potential biases.} 
\revised{Thus, while we find some promising support that interaction traces can promote conscious reflection of decision making strategies, additional studies are required to reach more conclusive results.}
\delete{Overall, we find that real-time and summative interaction trace techniques show promise toward promoting conscious reflection of one's own decision making.}

\acknowledgments{
This work was supported in part by the National Science Foundation
grant IIS-1813281 and the Siemens FutureMaker Fellowship. We thank the reviewers for their constructive feedback during the review phase. We also thank the Georgia Tech Visualization Lab for their feedback.}

\bibliographystyle{abbrv-doi}

\bibliography{template}
\end{document}